# Chiral topological whispering gallery modes formed by gyromagnetic photonic crystals

Yongqi Chen[1], Nan Gao[1], Guodong Zhu[1] and Yurui Fang[1, *]

[1.] *School of Physics, Dalian University of Technology, Dalian 116024, P.R. China.*
*Corresponding authors: yrfang@dlut.edu.cn (Y.F.)*

**Abstract:**

We explore a hexagonal cavity that supports chiral topological whispering gallery (CTWG) modes, formed by a gyromagnetic photonic crystal. This mode is a special type of topologically protected optical mode that can propagate in photonic crystals with chiral direction. Finite element method simulations show that discrete edge states exist in the topological band gap due to the coupling of chiral edge states and WG modes. Since the cavity only supports edge state modes with group velocity in only one direction, it can purely generate traveling modes and be immune to interference modes. In addition, we introduced defects and disorder to test the robustness of the cavity, demonstrating that the CTWG modes can be effectively maintained under all types of perturbations. Our topological cavity platform offers useful prototype of robust topological photonic devices. The existence of this mode can have important implications for the design and application of optical devices.

1. **Introduction**

   Topological photonics provides a novel exploration realm for manipulating the light-matter interaction through its protected topological photonic states in designing photonic crystals (PC) [1,2], becomes a flourishing field and a leading platform for the exploration of new phenomenon of nanophotonics [3–8]. Originated from condensed matter physics [9–11], photonic crystals analogous to "edge states" (ES) of a quantum Hall effect, topological photonics offers particularly exciting opportunities to enable robust propagation and resilience to noise and disorder. The special properties of topological photonics arise from their boundary topology, where the movement of light waves is restricted to certain paths known as ES, which can support robust light-wave propagation immune to backscattering and perturbations, having potential applications in elevating transmission quality and efficiency of optical signals. The boundary supports anti-direction propagation of ES with different pseudospin by considering the preserved time-reversal symmetry; and has been used in nanocavity, lasers, and so on [12–16]. Applying an external dc magnetic field will induce strong gyromagnetic anisotropy, which will break the time-reversal symmetry (TRS) so that the topological photonics will only support one-way propagation of light, known as chiral topological edge state (CES) [17,18]. Experimental demonstration of the CES in microwaves was performed using a magneto-optical photonic crystals (MOPC) immersed in an external DC magnetic field and advantageously employed in other a number of works [19,20]. The working frequency extends from microwave to terahertz and visible wavelength, which will contribute to future on-chip optical communications and ultrafast information processing. Currently, the CES with broken TRS inspires numerous novel phenomena, such as nonreciprocal laser in topological cavities with arbitrary geometric shapes [21], group-dispersionless slow light waveguides [22], and unidirectional guided waves [23–25].

   With the ingenious implementation of photonic crystal path-closed structures, the whispering gallery (WG) modes confining light beams has been realized in PC near closed geometric paths also have such advantages such as high-quality factors(Q) and small mode volumes(V) in many types of laser resonators, can significantly enhance the light-matter interaction [26–30]. The hybrid light-matter nature of PC provides a unique way to combine the WG modes and ES to study topological effects in lattices of resonators. The combination strategies will bring rich typical phenomena in topological isolator photonics. A honeycomb topological WG cavity was proposed based on the combination of topological PC and the geometric shape of WG modes cavities recently, which preserves TRS using the optical quantum spin Hall effect [31,32]. Compared with traditional WG cavities, the light confined to the edge of the topological WG cavity does not depend on total internal reflection but is a topological edge transport process. The laser nanocavity has high coherence and extremely low transmission loss, supports single-mode and robust optical transport, and has been applied to explore cavity quantum electrodynamics and high-performance lasers [33–37].

In above research where the system preserves TRS, the unidirectional transmission of light in the cavity significantly decreases with increasing lattice disorder. The size mismatch between the topological non-trivial lattice and the trivial lattice is required to construct the topological interface of the WG cavity, which breaks the lattice symmetry and makes the robustness of the cavity difficult to guarantee. In such system, the ES in WG is unidirectional modes but will be influenced at the corner of the cavity. Moreover, due to the presence of defects and corners, the two counter-propagation edge states supported by the system exhibit an increase in degeneracy, leading to interference and the creation of standing modes. In contrast, the above limitations will be minimized by a system that utilizes the broken TRS. Considering the properties of CES, one expects strictly one-way unidirectional WG modes by combining the CES and WG. The CES properties will guarantee the propagation of the light in WG is only in one direction. The successive advances of CESs in WG upon light propagation will have great potential in photonics and optoelectronics applications. The evolution from CES to WGMs will give us a comprehensive investigation of topological charges and topological properties in both theoretical and experimental investigations.

In this paper, we report topological WG cavity supporting only chiral nondegenerate traveling modes. With studying the transmission a hexagonal topological closed interface, we investigate the evolution of CES to chiral topological whispering gallery (CTWG) modes and demonstrate a hexagonal cavity that supports CTWG modes. The integrated CTWG cavity with precisely matched C6 symmetry, using a static magnetic field to break the time-reversal symmetry, allow only traveling modes to exist in the cavity. Next, using finite element methods (FEM), we show that discrete edge states can be observed in the topological band gap due to the coupling of topological edge states and WG modes. In addition, we effectively *forbid* the generation of standing modes in the cavity and test the robustness of the cavity by introducing defects and disorder. The results show that each mode in the cavity is traveling mode and that the coupling conditions between the CES and the cavity modes are still satisfied in the presence of disordered perturbations. The proposed CTWG cavity provides a potentially robust platform for constructing topological traveling wave generators, topological photonic circuits and topological laser at microwave frequencies, with enhanced fields-matter interactions. Its simple structural design provides a new research orientation for in-depth investigation of the interaction between photonic topological states and cavity modes. Given that its operating frequency is at communication band, this interaction is expected to trigger a wealth of physical phenomena and practical functions.

## 2. Results and discussion
### 2.1 Structure of chiral topological whispering gallery cavity

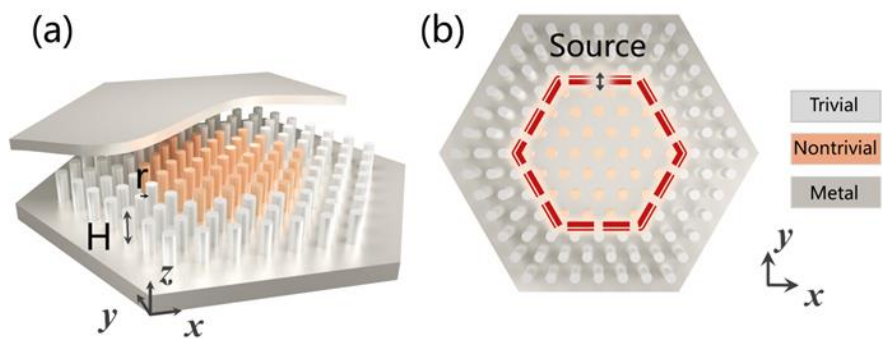

**Figure 1 Chiral topological whispering gallery cavity**. (**a**) The system is a three-layer structure, The middle layer is a topological whispering gallery cavity The top and bottom layers are parallel copper plates. (**b**) The top view of the topological whispering gallery cavity. The system is composed of two types of lattices: nontrivial and trivial unit cells highlighted in orange and silver, respectively. The red line marks the topological interface, in which a point source is located.

Similar to the non-trivial optical structure by combined WG cavity geometries with photonic topological insulators with TRS [38], Figure 1a illustrates the three-layer geometric structure of broken TRS CTWG cavity, with parallel gold plates on the top and bottom layers to confine electromagnetic waves in the z-direction. The middle layer is a CTWG cavity composed of triangular lattices of dielectric pillars PC, including trivial lattice surrounds topological lattice as shown in Figure 1b. The cell medium pillars (marked in silver) material of the peripheral is alumina with $\varepsilon=10$ in the air; and the surrounded pillars (marked in orange) is Yttrium Iron Garnet (YIG $\varepsilon=15$) MOPC, which is used to break the TRS of the system under static external magnetic field (EMF) with an externally applied static EMF $\mathbf{H}=H_0\mathbf{e}_z$ ( $H_0 = 1600$ G) are plot, where $\mathbf{e}_z$ is the unit vector perpendicular to the plane of the PC. The static EMF saturates the YIG PC material with spin-magnetic effects, resulting in strong anisotropy and maximum non-diagonal elements of the magnetic permeability tensor, which can be expressed as [39]:

$$\boldsymbol{\mu} = \begin{pmatrix} \mu & jk & 0 \\ -jk & \mu & 0 \\ 0 & 0 & 1 \end{pmatrix} \qquad (1)$$

where $\mu = 1 + \omega_m\omega_0/(\omega_0^2-\omega^2)$, $k = \omega_m\omega_0/(\omega_0^2-\omega^2)$ with $\omega_0 = \gamma H_0$ is the procession frequency, $\gamma$ is the gyromagnetic ratio, $H_0$ is the static EMF strength and $M_s$ in $\omega_m = 4\pi\gamma M_s$ describes the saturation magnetization strength.

The hexagonal topological interface formed by splicing the topological lattice and the trivial lattice creates a topological whispering gallery cavity with a perfectly matched C6 symmetry, as shown in the figure with a marked point source. The two PC have different topological invariants, ensuring the formation of robust CES at the topological interface, resulting in a CTWG cavity.

## 2.2 Band structures of topological nontrivial and trivial lattices

Here, topological indices can be used to distinguish topological non-trivial and trivial lattices. Based on the FEM provided by COMSOL Multiphysics [40], band structure of the two-dimensional (2D) PC in Figure 1 under *TM polarization* are calculated to reveal the topological properties of these two types of PC in the whispering gallery cavity. In order to construct the CTWG cavity structure, both the non-trivial and trivial photonic structures with *a=3.8 cm, r = 0.122a, H = a* triangular lattice PC slats are selected.

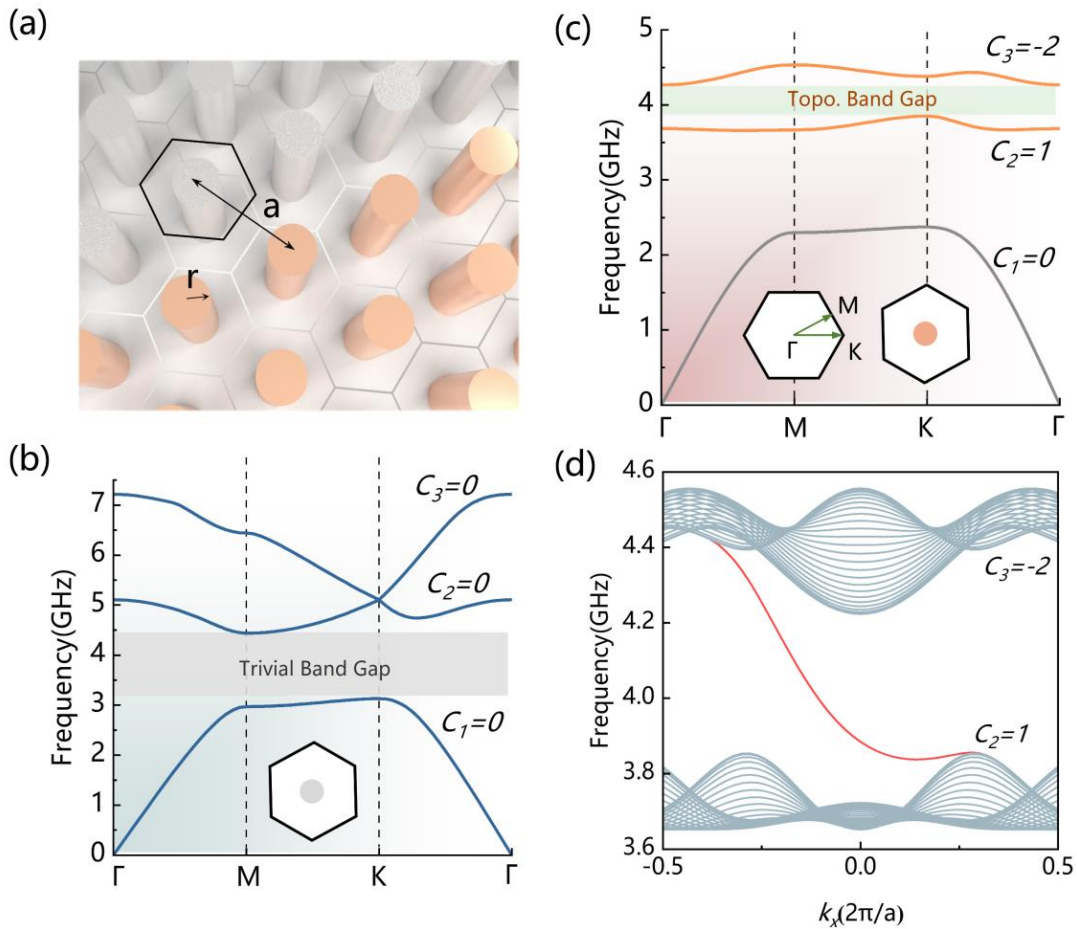

**Figure 2  Band structure and photonic properties of the topological interface.** (a) Detailed illustration of the topological interface, composed of nontrivial(orange) and trivial(silver) unit cells of the triangular lattice. they are composed of YIG magneto-optical and alumina PC respectively. In each unit cell, the lattice constant is *a=3.8 cm*, and the dielectric pillar radius is *r = 0.122a*. Photonic bands structure of trivial. (b) and nontrivial(c) photonic crystals, the insets illustrate the Brillouin zone and the k-path. (d) Projected band diagram. The dispersion curve (red) for the edge state spans the topological bandgap.

The details of the topological interface formed by the stitching of the two PC within the cavity are shown in Figure 2a. The band structures corresponding to the 2D PC are shown in Figure 2b-c, respectively. Figure 2b shows the band structure of the PC of alumina lattice. These crystals possess a triangular lattice that creates a broad bandgap within the desired frequency range which covers 3.2 to 4.5 GHz. In Figure 2c the band structure of the MOPC of the triangular lattice YIG pillar is plot with 2nd and 3rd band located in 3.9 to 4.3 GHz. In this band structure, the energy band is non-degenerated at K point. To investigate the topological properties of MOPC system of broken TRS, the topological invariants of a single band are described by the Chern number [41,42]:

$$C^{(n)} = \frac{1}{2\pi} \int_{BZ} \mathbf{F}_n(\mathbf{k}) d\mathbf{k}$$
$$= \frac{1}{2\pi} \int_{BZ} \nabla_\mathbf{k} \times \mathbf{A}_n(\mathbf{k}) d\mathbf{k} \quad (2)$$

Here, the $\mathbf{F}_n(\mathbf{k})$ is the Berry curvature. $\mathbf{A}_n(\mathbf{k}) = -i\langle u_{n,e,\mathbf{k}} | \nabla_\mathbf{k} | u_{n,e,\mathbf{k}} \rangle$ is the Berry connection. The Chern number of the $n^{th}$ band is an integer that can be calculated from the integral of the Berry curvature over all first Brillouin. We apply the calculation scheme of Jin et al. [20] to the calculation of the Chen number for the MOPC bands with a triangular lattice, for the discrete Brillouin zone as shown in Figure 3, and thus the Chern number can be further calculated by using:

$$C^{(n)} = \frac{1}{2\pi} \int_{BZ} \mathbf{F}_n(\mathbf{k}) d\mathbf{k}$$
$$= \frac{1}{2\pi i} \oint_{\partial BZ} \langle u_{n,e,\mathbf{k}} | \nabla_\mathbf{k} | u_{n,e,\mathbf{k}} \rangle d\mathbf{k}$$
$$= \frac{1}{2\pi} \sum_{\mathbf{k} \in BZ} F_\mathbf{k}^{(n)} \Delta S_\mathbf{k} \quad (3)$$
$$= \frac{1}{2\pi} \sum_{\mathbf{k} \in BZ} \operatorname{Im} \ln \left[ U_{\mathbf{k}_1 \to \mathbf{k}_2}^{(n)} U_{\mathbf{k}_2 \to \mathbf{k}_3}^{(n)} U_{\mathbf{k}_3 \to \mathbf{k}_4}^{(n)} U_{\mathbf{k}_4 \to \mathbf{k}_1}^{(n)} \right]$$

where $U_{\mathbf{k}_\alpha \to \mathbf{k}_\beta}^{(n)} \equiv \frac{\langle u_{n,e,\mathbf{k}_\alpha} | u_{n,e,\mathbf{k}_\beta} \rangle}{|\langle u_{n,e,\mathbf{k}_\alpha} | u_{n,e,\mathbf{k}_\beta} \rangle|}$ is a U(1) link variable, $|u_{n,e,\mathbf{k}}\rangle$ is the normalized eigenstate of the electric field, which satisfies the Bloch theorem [43,44]. And $\alpha, \beta = 1, 2, 3, 4$. $\mathbf{k}_1$, $\mathbf{k}_2$, $\mathbf{k}_3$, $\mathbf{k}_4$ are the vertices of any discrete Brillouin zone mesh (Figure 3b).

In the presence of an EMF, as depicted in Figure 2c, it is evident that the second and third bands exhibit a gap (indicated by the green-shaded region) within the desired frequency range. The sum of the Chern numbers corresponding to the bands below the band gap is $\left| \sum C_{gap} \right| = 1$, so the band gap with non-zero Chern numbers is a topological non-trivial band gap. This topological property is due to the TRS breaking of MOPC under an

EMF, which makes the medium have gyromagnetic anisotropy. The Chern numbers of alumina pillar PC in Figure 2b for the band below the gap are summed up to zero; and the band is considered as a trivial band gap.

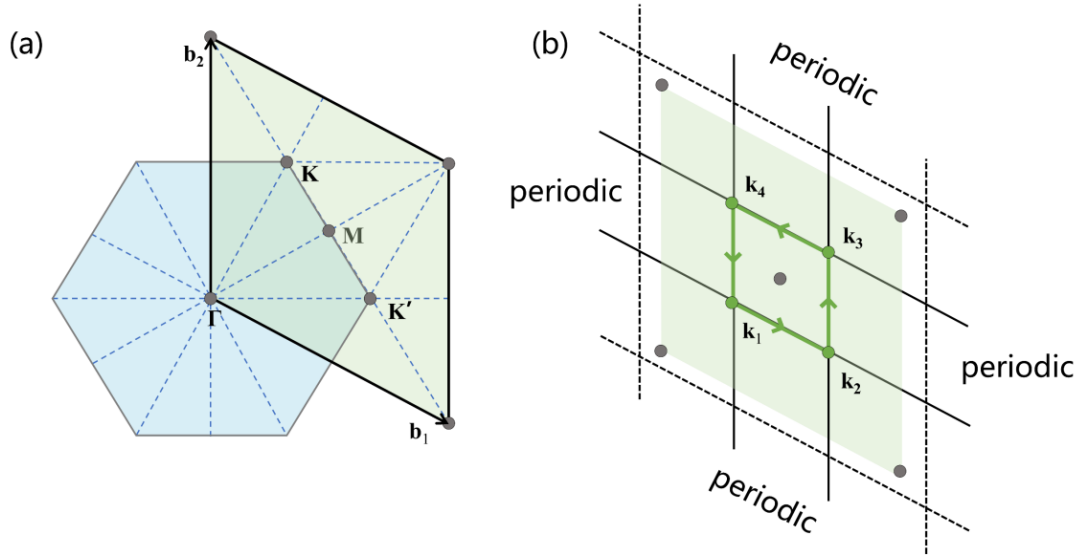

**Figure 3 Computational scheme for the calculation of Chern numbers.** **(a)** For the convenience of K-space discretization, the hexagonal Brillouin zone (blue region) is equivalently replaced by the rhombic Brillouin zone (green region) with high-symmetry points (**Γ**,**M**,**K**,and **K'**) indicated. **(b)** Discretized Brillouin zone, The line elements highlighted in green represent a Wilson loop that surrounds the **M** point.

In order to confirm the presence of CES at the topological boundary created by the two PC, a projected band diagram is generated through numerical simulation of the frequencies for the calculated topological non-trivial band gap, as illustrated in Figure 2d. This results in the observation of a CES dispersion curve with a non-zero group velocity (indicated by the red curve in Figure 2d across the topological bandgap). The direction of the group velocity of this edge state is dependent on the magnetic field direction in the vertical plane. As the bandgap possesses a non-zero Chern number, the edge state on this interface is protected by the topology. Experimental and theoretical simulations have demonstrated that this chiral topological edge state is robust against certain impurity defects and backscattering [45,46]. This advantage will be utilized in the construction of a whispering gallery cavity based on the chiral topological edge state.

**2.3 The topological whispering gallery modes based on chiral edge states.**

The existence of chiral of topological edge states will be established in the interface of above two PCs. The above size of the two PC is chosen to be within the near-infrared band, covering their respective band gaps, and they are defect-free stitched together to form the topological interface. A light source with a wavelength within the topological bandgap is positioned at the interface of two PC possessing distinct topological properties. When the static EMF direction is $+\mathbf{e_z}$, such as that shown in Figure 4a, the energy of the light source is significantly confined at the topological interface and is transmitted towards the right. However, when the direction of the EMF is reversed to $-\mathbf{e_z}$, such as that shown in Figure 4b, the wave propagating towards the right is prohibited. This observation confirms the presence of CES. The right side transmission spectra for two different EMF directions are presented in Figure 4c. When the magnetic field direction is $+\mathbf{e_z}$, a transmission peak appears at 4.03 GHz, and the entire transmission peak corresponds to the eigenfrequency of the non-trivial bandgap of the topological edge state. When reversing the magnetic field direction, however, the rightmost transmission disappears, and transmission to the right is forbidden, further demonstrating the existence of a chiral edge state.

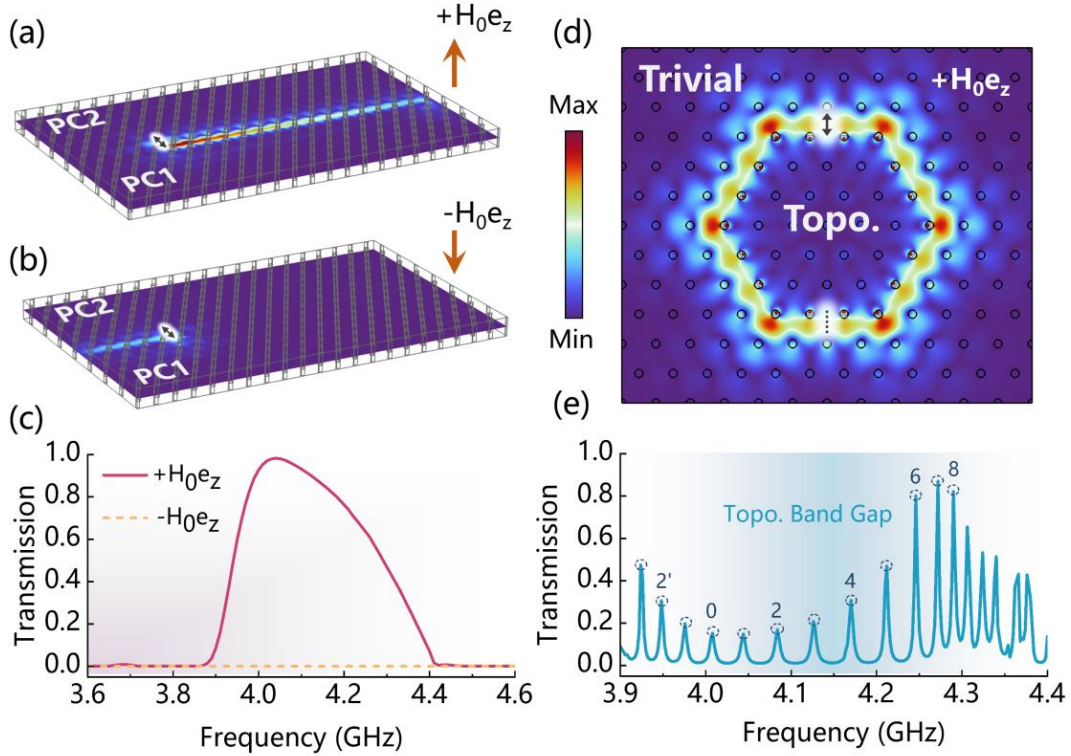

**Figure 4 Evolution of chiral edge states to CTWG modes.** The boundary between these two types of PCs with overlapping band gaps supports CES propagating to the right (**a**) or left (**b**), depending on the direction of the EMF. (**c**) The transmission spectrum shows the continuous peaks of the edge states for two opposite values of the EMF. (**d**) The eigenmode electric field distribution at 4.245 GHz of the CTWG modes with an EMF of $+H_0e_z$. The black double arrow indicates the source, and a dotted line draws the position of the detector. (**e**) The transmission spectra exhibit discrete peaks corresponding to the CTWG modes situated within the topological band gap, thereby signifying the transformation of the continuous CES into discrete CTWG modes.

The CTWG cavity can be conceptualized as a closed profile created by the topological interface after undergoing six 60° turns. It is worth noting that the cavity structure can theoretically take on any closed shape. Figure 4d illustrates the electric field distribution of the eigenmode at 4.245 GHz within the hexagonal whispering gallery, with the excitation frequency of the light source situated within the topological band gap. In the same platform as the above illustration (Figure 1), the side length of the whispering gallery cavity is 4a and the internal topological non-trivial photonic crystal is surrounded by the trivial photonic crystals. The black double arrows and the dotted line in the figure indicate the positions of the source and the detector, respectively. Numerical simulations indicate that the impact of light propagation through a 60° turn angle is negligible, and the field energy is significantly concentrated within the whispering gallery cavity. The scattering of the corner defect is avoidable in the CTWG. When the EMF is oriented in the $+\mathbf{e}_z$ direction, the CES, which initially propagates towards the right, transforms into a clockwise mode within the whispering gallery cavity. These cavity modes arise because of coupling of the cavity and the CES propagating around the cavity. To demonstrate this property, the same point source utilized in Figure 4a was positioned within the CES CTWG cavity, and the transmission spectrum to symmetric position of the other side of the hexagon was computed, as depicted in Figure 4e. When a topological edge state is coupled into a whispering gallery cavity, its transmission exhibits a distinctive interference effect that produces multiple sharp peaks. Each peak corresponds to a topological whispering gallery mode, and the emergence of discrete CTWG modes is a distinctive feature of the CES interaction with the cavity mode. The presence of multiple CTWG modes can be attributed to constructive interference in the CES during clockwise transport within the cavity,

$$k_{//}(\omega)\Lambda = 2m\pi \tag{4}$$

where is the wave number $k_{//}$ corresponding to the CTWG modes, $\Lambda$ is the circumference of the cavity, and $m$ corresponds to the number of azimuthal modes.

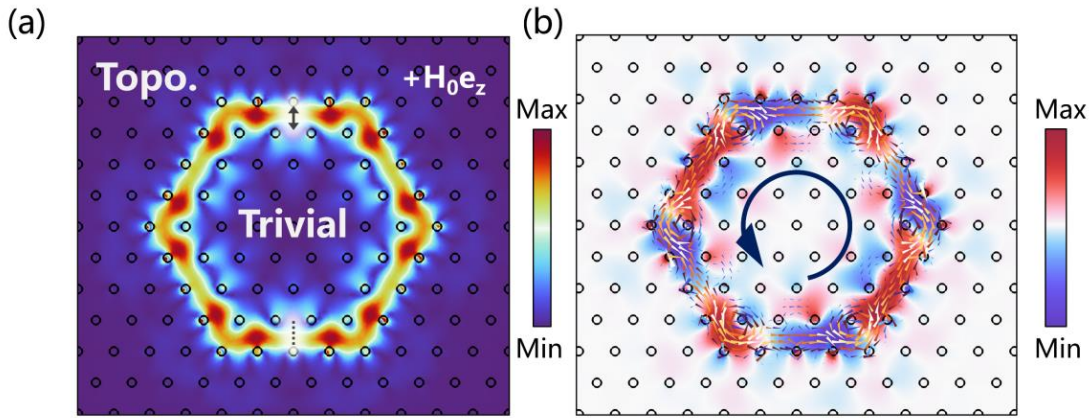

**Figure 5 Topological indices reverse of CTWG cavity. (a)** The eigenmode electric field distribution at 4.0906 GHz of the CTWG modes for an EMF of $+H_0e_z$. Compared to Figure 4d, when the internal trivial photonic crystals are surrounded by topological non-trivial photonic crystals, resulting in a counterclockwise mode **(b)** within the whispering gallery cavity. The arrows on the CTWG cavity represent the Poynting vector distribution for the *4th* order mode, and the colors show the electric field mode intensity.

To further demonstrate that traveling modes in topological whispering gallery cavity are chiral, we reverse the structure on both sides of the topological interface, where the internal trivial PCs are surrounded by topological non-trivial PCs. Figure 5a illustrates the electric field distribution of the eigenmode after reversal, where a cavity mode localized at the interface between the two PCs was observed in the presence of the EMF. Further mapping of the distribution for the time-averaged Poynting vectors in the topological whispering gallery cavity exhibits evident that the reversed structure results in a counterclockwise mode, as shown in Figure 5b. It is worth noting that all modes within this CTWG cavity are nondegenerate modes due to the broken TRS within the system, which distinguishes them from conventional whispering gallery modes. Especially, the unique cavity modes are all one-way unidirectional which originate from the CES that is the term CTWG modes.

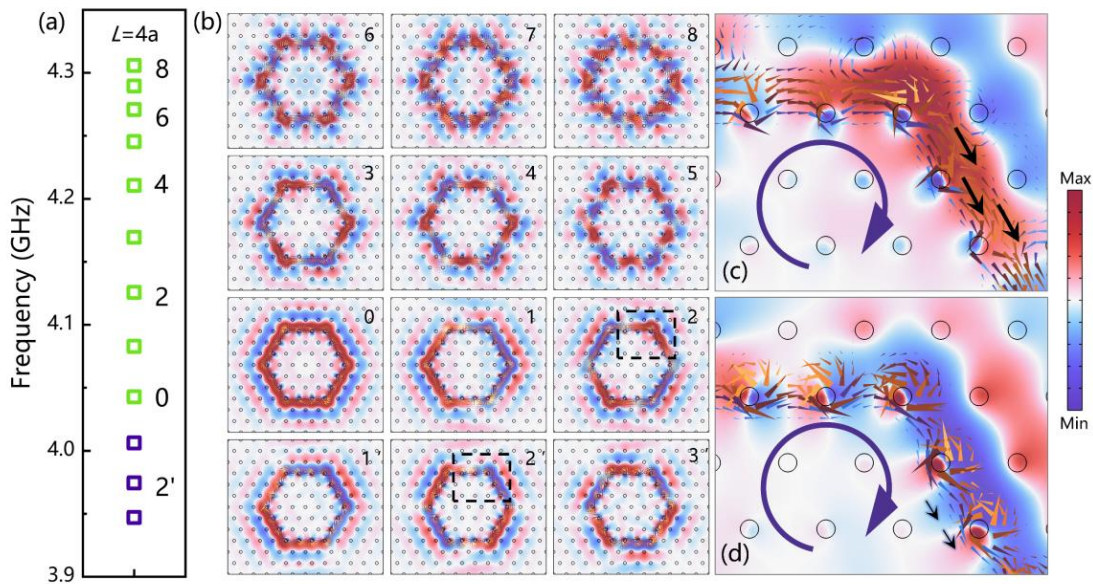

**Figure 6 Mode analysis of CTWG cavity. (a)** Frequency spectrum of CTWG cavity with side length *L = 4a* contains twelve CTWG modes. The discrete CTWG modes labeled by the number of azimuthal modes. **(b)** $E_z$ field distributions for all the CTWG modes are shown, the colors red and blue are used to indicate the relative phase of a field. **(c)-(d)** Enlarged view of the 2$^{nd}$ **(c)** and 2$'^{nd}$ **(d)** clockwise modes, where the arrows indicate the distribution of time-averaged Poynting vector for the CTWG mode.

The CTWG cavity in Figure 4d is used as an example to analyze its eigenmodes. The spectrum of the cavity is obtained when the wave vector $\mathbf{k}=0$, as shown in Figure 6a. Within the topological band gap, a total of twelve nondegenerate modes are marked (bulk states not shown), which evolve from continuous topological edge states into the discrete modes. Figure 6b displays the $E_z$ field distributions of all of these eigenmodes, where one can observe alternating red and blue patterns on the azimuth of the CTWG cavity, also corresponding to the propagation field phase varying of the cavity mode. The $0^{th}$ mode is marked as $0^{th}$ in the absence of an azimuthal node, and the corresponding marker for each eigenmode matches the number of CTWG azimuthal modes. The modes above (green dot) and below (purple dot) the $0^{th}$ CTWG mode correspond to two orthogonal modes states, the high frequency state and the low-frequency state, respectively. It should be emphasized that the modes in a conventional whispering gallery cavity can be classified into two categories: standing modes and traveling modes [47]. However, despite the CTWG cavity having C6 symmetry, all modes in this cavity are traveling modes due to the forbidding counter-propagating modes. To confirm this, Figure 6c and d displays the Poynting vector distribution for the second-order mode, revealing that the Poynting vectors for both the $2^{nd}$ (eigenfrequency $\omega_2$=4.122 GHz ) and $2'^{nd}$ CTWG modes (eigenfrequency $\omega_{2'}$=3.974 GHz) exhibit a clockwise flow characteristic along the cavity, which is consistent with the direction of the CES shown in Figure 4a. Furthermore, it is worth noting that the $2^{nd}$ CTWG mode (high-frequency state) exhibits a greater concentration of clockwise mode fields in topological interface band between the trivial and non-trivial pillars compared to the $2'^{nd}$ CTWG mode (lower-frequency state). In contrast, the field energy in the low-frequency state is concentrated around the MOPC dielectric pillars within the non-trivial side. As per the energy variational principle [48], it is well-established that the cavity mode tends to concentrate the majority of its electric field energy in the high dielectric constant region to minimize its frequency. Moreover, it is also important to note that the upper-frequency band must be orthogonal to the lower-frequency band. The $2'^{nd}$ mode achieves lower frequencies by concentrating the clockwise mode on the inner side of the interface. In contrast, the $2^{nd}$ mode, which is symmetric to the low-frequency state, shifts the clockwise mode towards the trivial photonic crystal and into the topological interface to maintain orthogonality with the $2'^{nd}$ mode. This is the reason why the $7^{th}$ CTWG mode exhibits a high transmission peak, while the $8^{th}$ mode has reduced transmission due to its frequency being outside the topological band gap (Figure 4e).

In the case of this cavity, the number of CTWG modes increases as the length of the cavity increases. As we increase the size of the cavity, additional modes move from the bulk band to the edge state, as demonstrated in Eq. 4. As the cavity length increases from *4a* to *5a*, the number of CTWG modes, shown in Figure 7a, increases from twelve to sixteen.

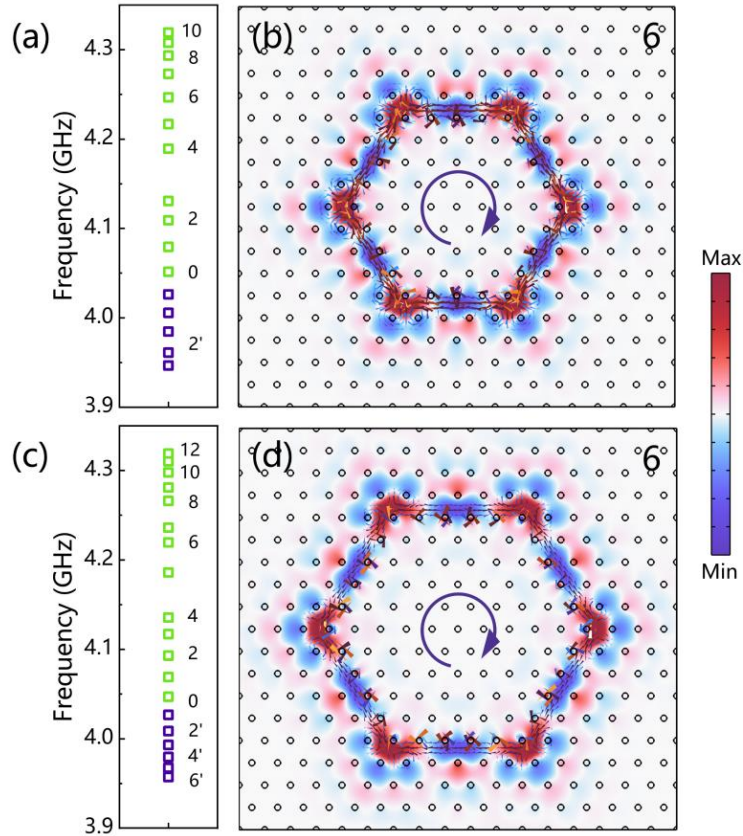

**Figure 7 CTWG mode distribution at different cavity lengths.** The discrete CTWG modes labeled by numbers calculated with *L = 5a* (**a**), and *L = 6a* (**b**) respectively. (**c**), (**d**) The 6$^{th}$ $E_z$ field and Poynting vector distribution of the corresponding cavity length show a clockwise traveling mode.

Similarly, for a cavity with an edge length of *6a*, the band gap contains nineteen modes, as depicted in the Figure 7c. Even if the length of the cavity is increased, the CTWG modes still maintains the clockwise traveling wave characteristics, as shown in Figures 7c-d (taking the sixth-order mode as an example). However, for cavities smaller than *4a*, the CTWG modes will disappear because there is not enough bulk to support the generation of edge states unless the lattice constant becomes even smaller. Therefore, the number of CTWG modes is completely geometrically dependent. This effect has been predicted in photonic topological insulator strips [49].

**2.4 Robustness of CTWG modes**

Robustness refers to the ability of a system to maintain its stability and reliability in the face of external interference or internal changes. The ability to maintain robustness against certain types of disorder under necessary symmetry protection is a remarkable feature of topological photonic systems. To investigate the topological protection properties of this CTWG cavity modes, we considered an array of cavities with length of *4a* and randomly varied the magnitudes of the dielectric pillars radius of the non-trivial unit cells, isotropically and anisotropically. The eigenspectrum of the cavity are subsequently calculated in the presence of pillar radius disorder, as shown in Figure 8a with the inset. The results showed that the CTWG mode remained stably present within the topological band gap, which is consistent with the findings in Figure 6a.

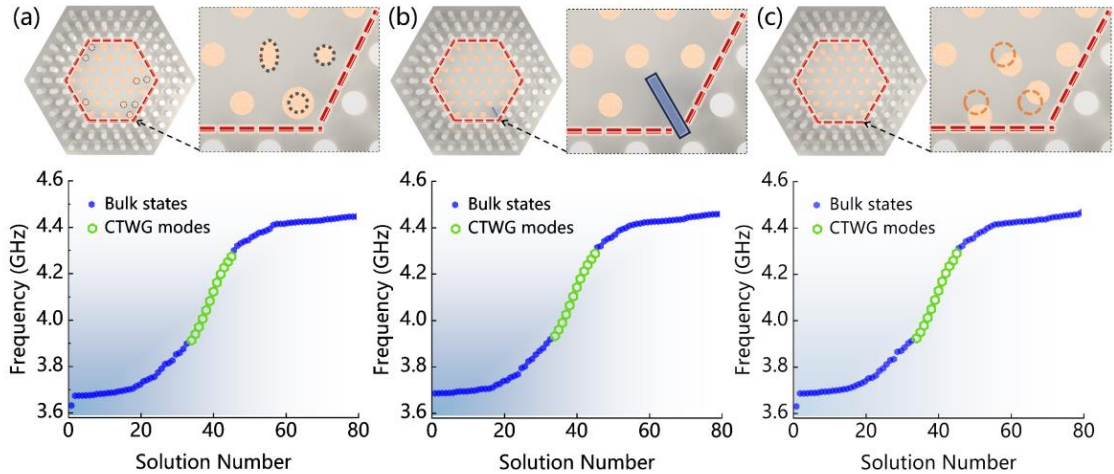

**Figure 8** Topological protection characteristics of whispering gallery modes. (**a**) Eignspectra of the CTWG cavity with the disorders of dielectric pillars size. The inset indicates the location of the disordered unit cells. (**b**) Eigenspectra of CTWG cavity containing defects and contaminants, the blue areas indicate impurities at defects. (**c**) Eigenspectra of CTWG cavity under positional disorder.

Next, a strong symmetry-breaking perturbation was induced by introducing defects and impurities in the cavity. The eigenspectrum of the CTWG modes under this condition was plotted in Figure 8b, and it was observed that the modes remained robustness. Figure 8c presents the eigenspectrum of the cavity when there is a certain amount of disorder at the unit cell location. It was found that the spectral profile line shape can be matched to the line shape when it is unperturbed and supports the same number of CTWG modes. Hence, the topologically protected coupling between the edge states and the whispering gallery modes persists despite any potential perturbations. This affirms the topological protection of the CTWG modes to a considerable extent.

3. **Conclusion**

In summary, we have constructed a CTWG cavity by using topological non-trivial photonic crystals surrounded by trivial PC, where each mode is a one-way unidirectional traveling mode. We investigated the CES energy band structure and topological properties. Then a comprehensive investigation on the coupling characteristics between the CES and the whispering gallery cavity in the near-infrared range was performed, which reveals that the originally continuous topological edge states evolved into multiple discrete nondegenerate whispering gallery modes in the cavity, with the number of modes only related to the circumference of the topological whispering gallery cavity. The CES guarantees that the wave can only propagate in one-way, and thus in the CTWG cavity, there are only traveling modes existing. The high-frequency states concentrated in topological interface band has extremely high transmission efficiency. which differs from whispering gallery cavities that support both travelling and standing waves We further investigate the robustness of the CTWG modes derived from topological protection. Numerical simulations showed that the coupling of edge states and cavities and the cavity modes remained robust even in the presence of significant disorder. This means that the CTWG cavity is able to maintain stability and reliability in the face of external interference or internal changes, making it a promising candidate for various applications, allowing for the discovery and implementation of fruitful physical phenomena and useful functionalities.

**Acknowledgements**

The authors thank the supporting of the National Natural Science Foundation of China (Grant No. 12074054, 12274054).

**Author contribution**

Y.F. conceived the idea and directed the project. Y.C. performed the FEM calculations. Y.C. and Y.F analyzed the data and wrote the manuscript. All the authors revised the manuscript.

**Conflicts of interest**

The authors declare no competing financial interest.

**Methods of FEM Simulation**

Numerical simulations are performed using a commercial finite-element simulation software (COMSOL MULTIPHYSICS) via the Optical module. The band structures are simulated for the 2D transverse magnetic mode (i.e., the electric field is perpendicular to the 2D plane, along the z-direction). The projected band diagram is calculated by simulating the eigenfrequencies of a semi-infinite topological non-trivial lattice. While the transmission spectra of the CES and CTWG modes are calculated in 3D photonic crystals with 21×8 and 12 unit cells depicted in Fig.3. Mode analysis in Figure 6 is performed on the 3D CTWG cavity and the wave vector is set as $k=0$.

**References**

[1] S. Raghu and F. D. M. Haldane, *Analogs of Quantum-Hall-Effect Edge States in Photonic Crystals*, Phys. Rev. A **78**, 3 (2008).
[2] L. Lu, J. D. Joannopoulos, and M. Soljačić, *Topological Photonics*, Nat. Photonics **8**, 11 (2014).
[3] M. S. Rider, S. J. Palmer, S. R. Pocock, X. Xiao, P. Arroyo Huidobro, and V. Giannini, *A Perspective on Topological Nanophotonics: Current Status and Future Challenges*, J. Appl. Phys. **125**, 12 (2019).
[4] F. Liu, H.-Y. Deng, and K. Wakabayashi, *Topological Photonic Crystals with Zero Berry Curvature*, Phys. Rev. B **97**, 3 (2018).
[5] Z. Zhang, J. Yang, T. Du, and X. Jiang, *Topological Multipolar Corner State in a Supercell Metasurface and Its Interplay with Two-Dimensional Materials*, Photonics Res. **10**, 4 (2022).
[6] L.-H. Wu and X. Hu, *Scheme for Achieving a Topological Photonic Crystal by Using Dielectric Material*, Phys. Rev. Lett. **114**, 22 (2015).
[7] S. Han et al., *Electrically-Pumped Compact Topological Bulk Lasers Driven by Band-Inverted Bound States in the Continuum*, Light Sci. Appl. **12**, 1 (2023).
[8] B. Hu et al., *Non-Hermitian Topological Whispering Gallery*, Nature **597**, 7878 (2021).
[9] M. Z. Hasan and C. L. Kane, *Colloquium: Topological Insulators*, Rev. Mod. Phys. **82**, 4 (2010).
[10] X.-L. Qi and S.-C. Zhang, *Topological Insulators and Superconductors*, Rev. Mod. Phys. **83**, 4 (2011).
[11] B. A. Bernevig and S.-C. Zhang, *Quantum Spin Hall Effect*, Phys. Rev. Lett. **96**, 10 (2006).
[12] M. A. Bandres, S. Wittek, G. Harari, M. Parto, J. Ren, M. Segev, D. N. Christodoulides, and M. Khajavikhan, *Topological Insulator Laser: Experiments*, Science **359**, 6381 (2018).
[13] G. Siroki, P. A. Huidobro, and V. Giannini, *Topological Photonics: From Crystals to Particles*, Phys. Rev. B **96**, 4 (2017).
[14] Y. Chen, X.-T. He, Y.-J. Cheng, H.-Y. Qiu, L.-T. Feng, M. Zhang, D.-X. Dai, G.-C. Guo, J.-W. Dong, and X.-F. Ren, *Topologically Protected Valley-Dependent Quantum Photonic Circuits*, Phys. Rev. Lett. **126**, 23 (2021).
[15] Y. Zeng et al., *Electrically Pumped Topological Laser with Valley Edge Modes*, Nature **578**, 246 (2020).
[16] X. Wu, Y. Meng, J. Tian, Y. Huang, H. Xiang, D. Han, and W. Wen, *Direct Observation of Valley-Polarized Topological Edge States in Designer Surface Plasmon Crystals*, Nat. Commun. **8**, 1 (2017).
[17] L. Esaki, *Highlights in Condensed Matter Physics and Future Prospects* (Springer Science & Business Media, 2013).
[18] Z. Wang, Y. D. Chong, J. D. Joannopoulos, and M. Soljačić, *Reflection-Free One-Way Edge Modes in a Gyromagnetic Photonic Crystal*, Phys. Rev. Lett. **100**, 1 (2008).
[19] Z. Wang, Y. Chong, J. D. Joannopoulos, and M. Soljačić, *Observation of Unidirectional Backscattering-Immune Topological Electromagnetic States*, Nature **461**, 7265 (2009).

[20] D. Jin, T. Christensen, M. Soljačić, N. X. Fang, L. Lu, and X. Zhang, *Infrared Topological Plasmons in Graphene*, Phys. Rev. Lett. **118**, 245301 (2017).

[21] Babak Bahari, A. Ndao, F. Vallini, A. El Amili, Y. Fainman, and B. Kanté, *Nonreciprocal Lasing in Topological Cavities of Arbitrary Geometries*, Science **358**, 6363 (2017).

[22] J. Chen, W. Liang, and Z.-Y. Li, *Strong Coupling of Topological Edge States Enabling Group-Dispersionless Slow Light in Magneto-Optical Photonic Crystals*, Phys. Rev. B **99**, 1 (2019).

[23] K. Fang, Z. Yu, and S. Fan, *Realizing Effective Magnetic Field for Photons by Controlling the Phase of Dynamic Modulation*, Nat. Photonics **6**, 11 (2012).

[24] C. He, X.-C. Sun, X.-P. Liu, M.-H. Lu, Y. Chen, L. Feng, and Y.-F. Chen, *Photonic Topological Insulator with Broken Time-Reversal Symmetry*, Proc. Natl. Acad. Sci. **113**, 18 (2016).

[25] L. Lu, H. Gao, and Z. Wang, *Topological One-Way Fiber of Second Chern Number*, Nat. Commun. **9**, 1 (2018).

[26] S. Yang, Y. Wang, and H. Sun, *Advances and Prospects for Whispering Gallery Mode Microcavities*, Adv. Opt. Mater. **3**, 9 (2015).

[27] K. J. Vahala, *Optical Microcavities*, Nature **424**, 6950 (2003).

[28] Y. Zhi, X.-C. Yu, Q. Gong, L. Yang, and Y.-F. Xiao, *Single Nanoparticle Detection Using Optical Microcavities*, Adv. Mater. **29**, 12 (2017).

[29] J. Wiersig, *Structure of Whispering-Gallery Modes in Optical Microdisks Perturbed by Nanoparticles*, Phys. Rev. A **84**, 6 (2011).

[30] J. A. Haigh, S. Langenfeld, N. J. Lambert, J. J. Baumberg, A. J. Ramsay, A. Nunnenkamp, and A. J. Ferguson, *Magneto-Optical Coupling in Whispering-Gallery-Mode Resonators*, Phys. Rev. A **92**, 6 (2015).

[31] Y. Yang and Z. H. Hang, *Topological Whispering Gallery Modes in Two-Dimensional Photonic Crystal Cavities*, Opt. Express **26**, 21235 (2018).

[32] L. He, W. X. Zhang, and X. D. Zhang, *Topological All-Optical Logic Gates Based on Two-Dimensional Photonic Crystals*, Opt. Express **27**, 25841 (2019).

[33] A. Dikopoltsev et al., *Topological Insulator Vertical-Cavity Laser Array*, Science **373**, 1514 (2021).

[34] M. A. Bandres, S. Wittek, G. Harari, M. Parto, J. Ren, M. Segev, D. N. Christodoulides, and M. Khajavikhan, *Topological Insulator Laser: Experiments*, Science **359**, eaar4005 (2018).

[35] M.-S. Wei, M.-J. Liao, C. Wang, C. Zhu, Y. Yang, and J. Xu, *Topological Laser with Higher-Order Corner States in the 2-Dimensional Su-Schrieffer-Heeger Model*, Opt. Express **31**, 3427 (2023).

[36] Y. Gong, S. Wong, A. J. Bennett, D. L. Huffaker, and S. S. Oh, *Topological Insulator Laser Using Valley-Hall Photonic Crystals*, ACS Photonics **7**, 2089 (2020).

[37] Y. Ota, R. Katsumi, K. Watanabe, S. Iwamoto, and Y. Arakawa, *Topological Photonic Crystal Nanocavity Laser*, Commun. Phys. **1**, 1 (2018).

[38] X.-C. Sun and X. Hu, *Topological Ring-Cavity Laser Formed by Honeycomb Photonic Crystals*, Phys. Rev. B **103**, 24 (2021).

[39] Z. Wang, Y. D. Chong, J. D. Joannopoulos, and M. Soljačić, *Reflection-Free One-Way Edge Modes in a Gyromagnetic Photonic Crystal*, Phys. Rev. Lett. **100**, 013905 (2008).

[40] *COMSOL - Software for Multiphysics Simulation*, https://www.comsol.com/.

[41] S. A. Skirlo, L. Lu, and M. Soljačić, *Multimode One-Way Waveguides of Large Chern Numbers*, Phys. Rev. Lett. **113**, 113904 (2014).

[42] S. A. Skirlo, L. Lu, Y. Igarashi, Q. Yan, J. Joannopoulos, and M. Soljačić, *Experimental Observation of Large Chern Numbers in Photonic Crystals*, Phys. Rev. Lett. **115**, 253901 (2015).

[43] T. Fukui, Y. Hatsugai, and H. Suzuki, *Chern Numbers in Discretized Brillouin Zone: Efficient Method of Computing (Spin) Hall Conductances*, J. Phys. Soc. Jpn. **74**, 1674 (2005).

[44] R. Zhao, G.-D. Xie, M. L. N. Chen, Z. Lan, Z. Huang, and W. E. I. Sha, *First-Principle Calculation of Chern Number in Gyrotropic Photonic Crystals*, Opt. Express **28**, 4638 (2020).

[45] Y. Poo, R. Wu, Z. Lin, Y. Yang, and C. T. Chan, *Experimental Realization of Self-Guiding Unidirectional Electromagnetic Edge States*, Phys. Rev. Lett. **106**, 093903 (2011).

[46] X. Ao, Z. Lin, and C. T. Chan, *One-Way Edge Mode in a Magneto-Optical Honeycomb Photonic Crystal*, Phys. Rev. B **80**, 033105 (2009).

[47] Q. Li, T. Wang, Y. Su, M. Yan, and M. Qiu, *Coupled Mode Theory Analysis of Mode-Splitting in Coupled Cavity System*, Opt. Express **18**, 8367 (2010).

[48] J. D. Joannopoulos, editor , *Photonic Crystals: Molding the Flow of Light*, 2nd ed (Princeton University Press, Princeton, 2008).

[49] L.-H. Wu and X. Hu, *Scheme for Achieving a Topological Photonic Crystal by Using Dielectric Material*, Phys. Rev. Lett. **114**, 223901 (2015).